\documentclass[conference]{IEEEtran}
\usepackage{acronym}
\usepackage{cite}
\usepackage{graphicx}
\graphicspath{{graphics/}}

\usepackage{url}
\usepackage{color}

\usepackage[colorinlistoftodos,prependcaption,textsize=small]{todonotes}
\usepackage{blindtext}
\presetkeys%
    {todonotes}%
    {inline,backgroundcolor=yellow}{}

\begin{document}
\title{Fault-Tolerant Adaptive Parallel and Distributed Simulation\footnotemark}

\author{\IEEEauthorblockN{Gabriele D'Angelo \quad Stefano Ferretti \quad Moreno Marzolla}
\IEEEauthorblockA{Dept. of Computer Science and Engineering, University of Bologna, Italy\\
Email: \{g.dangelo,s.ferretti,moreno.marzolla\}@unibo.it}
\and
\IEEEauthorblockN{Lorenzo Armaroli}
\IEEEauthorblockA{Email: lorenzo.armaroli@gmail.com}
}

\maketitle

\footnotetext{The publisher version of this paper is available at \url{https://doi.org/10.1109/DS-RT.2016.11}.
\textbf{{\color{red}Please cite this paper as: ``Gabriele D'Angelo, Stefano Ferretti, Moreno Marzolla, Lorenzo Armaroli. Fault-Tolerant Adaptive Parallel and Distributed
Simulation. Proceedings of the IEEE/ACM International Symposium on Distributed Simulation and Real Time Applications (DS-RT 2016)''.}}}

\begin{abstract}
  Discrete Event Simulation is a widely used technique that is used to
  model and analyze complex systems in many fields of science and
  engineering. The increasingly large size of simulation models poses
  a serious computational challenge, since the time needed to run a
  simulation can be prohibitively large. For this reason, Parallel and
  Distributes Simulation techniques have been proposed to take
  advantage of multiple execution units which are found in multicore
  processors, cluster of workstations or HPC systems. The current
  generation of HPC systems includes hundreds of thousands of
  computing nodes and a vast amount of ancillary components. Despite
  improvements in manufacturing processes, failures of some components
  are frequent, and the situation will get worse as larger systems are
  built. In this paper we describe FT-GAIA, a software-based
  fault-tolerant extension of the GAIA/ART\`IS parallel simulation
  middleware. FT-GAIA transparently replicates simulation entities and
  distributes them on multiple execution nodes. This allows the
  simulation to tolerate crash-failures of computing nodes;
  furthermore, FT-GAIA offers some protection against byzantine
  failures since synchronization messages are replicated as well, so
  that the receiving entity can identify and discard corrupted
  messages. We provide an experimental evaluation of FT-GAIA on a
  running prototype. Results show that a high degree of fault tolerance
  can be achieved, at the cost of a moderate increase in the
  computational load of the execution units.
\end{abstract}

\IEEEpeerreviewmaketitle

\section{Introduction}\label{sec:introduction}

Computer-assisted modeling and simulation plays an important role in
many scientific disciplines: computer simulations help to understand
physical, biological and social phenomena. \ac{DES} is of particular
interest, since it is frequently employed to model and analyze many
types of systems, including computer architectures, communication
networks, street traffic, and others.

In a~\ac{DES}, the system is described as a set of interacting
entities; the state of the simulator is updated by simulation
\emph{events}, which happen at discrete points in time. The overall
structure of a sequential event-based simulator is relatively simple:
the simulator engine maintains a list, called~\ac{FEL}, of all pending
events, sorted in non decreasing time of occurrence. The simulator
executes a loop, where at each iteration, the event with lower
timestamp~$t$ is removed from the~\ac{FEL}, and the simulation time is
advanced to~$t$. Then, the event is executed, possibly triggering the
generation of new events to be scheduled for execution at some future
time.

\begin{figure}[t]
  \centering\includegraphics[width=.8\columnwidth]{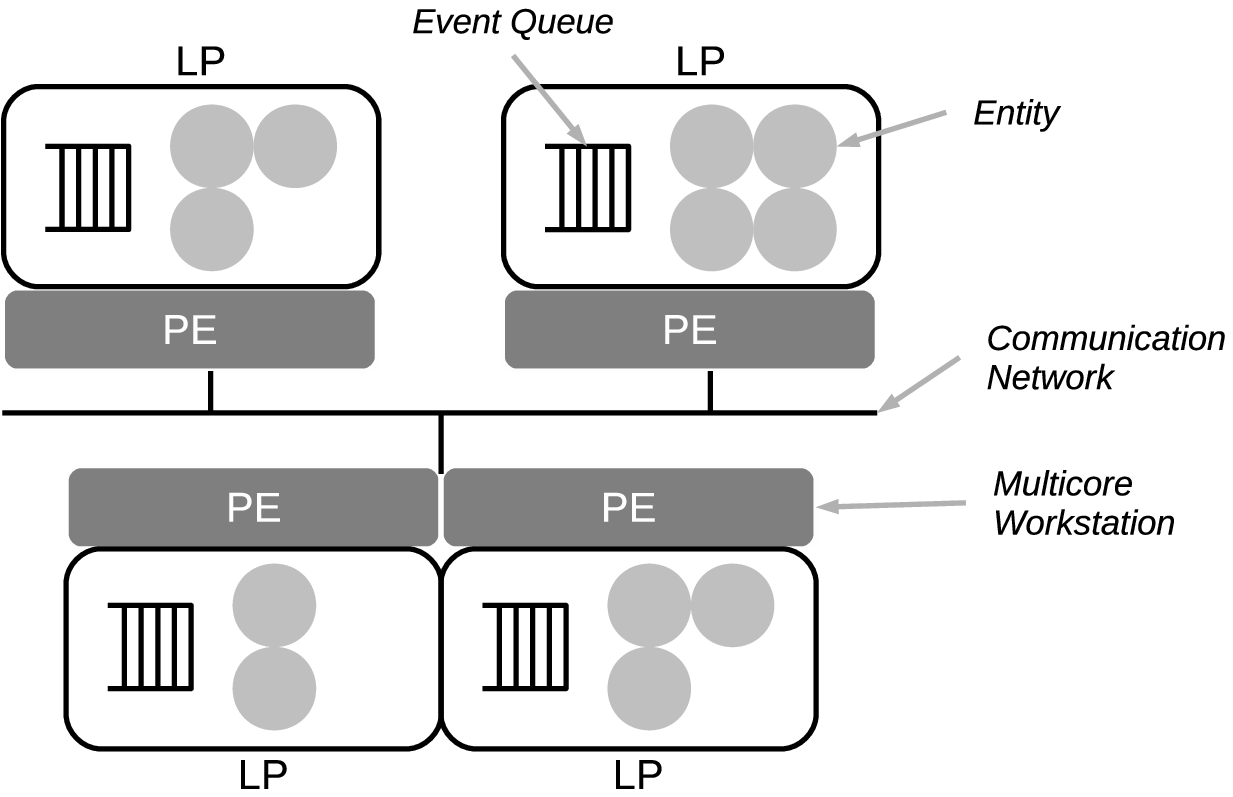}
  \caption{Structure of a~\acl{PADS}.}\label{fig:pads}
\end{figure}

Continuous advances in our understanding of complex systems, combined
with the need for higher model accuracy, demand an increasing amount
of computational power and represent a major challenge for the
capabilities of the current generation of high performance computing
systems. Therefore, sequential~\ac{DES} techniques may be
inappropriate for analyzing large or detailed models, due to the huge
number of events that must be processed. \ac{PADS} aims at taking
advantage of modern high performance computing architectures -- from
massively parallel computers to multicore processors -- to handle
large models efficiently~\cite{Fuj00}. The general idea of~\ac{PADS}
is to partition the simulation model into submodels, called~\acp{LP}
which can be evaluated concurrently by different~\acp{PE}. More
precisely, the simulation model is described in terms of multiple
interacting~\acp{SE} which are assigned to
different~\acp{LP}. Each~\ac{LP} is executed on a different~\ac{PE},
and is in practice the container of a set of entities. The execution
of the simulation is obtained through the exchange of timestamped
messages (representing simulation events) between
entities. Each~\ac{LP} has an queue where messages are inserted before
being dispatched to the appropriate entities. Figure~\ref{fig:pads}
shows the general structure of a parallel and distributed simulator.

Execution of long-running applications on increasingly larger parallel
machines is likely to hit the \emph{reliability
  wall}~\cite{reliability-wall}. This means that, as the system size
(number of components) increases, so does the probability that at
least one of those components fails, therefore reducing the
system~\ac{MTTF}. At some point the execution time of the parallel
application may become larger than the~\ac{MTTF} of its execution
environment, so that the application has little chance to terminate
normally.

As a purely illustrative example, let us consider a parallel machine
with $N$~\acp{PE}. Let $X_i$ be the stochastic variable representing
the duration of uninterrupted operation of the $i$-th~\ac{PE}, taking
into account both hardware and software failures. Assuming that all
$X_i$ are independent and exponentially distributed (this assumption
is somewhat unrealistic but widely used~\cite{bolch}), we have that
the probability $P(X_i > t)$ that~\ac{LP} $i$ operates without
failures for at least $t$ time units is
\[
P(X_i > t) = e^{-\lambda t}
\]
\noindent where $\lambda$ is the failure rate. The joint probability
that all $N$~\acp{LP} operate without failures for at least $t$ time
units is therefore $R(N, t) = \prod_i P(X_i > t) = e^{-N \lambda t}$;
this is the formula for the reliability of $N$ components connected in
series, where each component fails independently, and a single failure
brings down the whole system.

\begin{figure}[t]
  \centering%
  \includegraphics{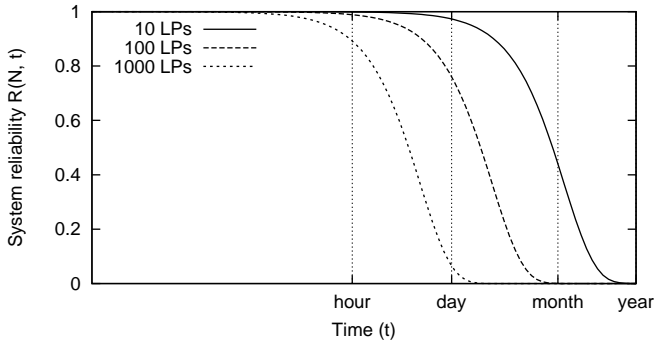}
  \caption{System reliability $R(N, t)$ assuming a MTTF for
    each~\ac{LP} of one year; higher is better, log scale on the $x$
    axis.}\label{fig:reliability}
\end{figure}

Figure~\ref{fig:reliability} shows the value of $R(N, t)$ (the
probability of no failures for at least $t$ consecutive time units)
for systems with $N=10, 100, 1000$ \acp{LP}, assuming a~\ac{MTTF} of
one year ($\lambda \approx 2.7573 \times 10^{-8} s^{-1}$). We can see
that the system reliability quickly drops as the number of~\acp{LP}
increases: a simulation involving $N=1000$ \acp{LP} and requiring one
day to complete is very unlikely to terminate successfully. 

Although the model above is overly simplified, and is not intended to
provide an accurate estimate of the reliability of actual parallel
simulations, it does show that building a reliable system out of a
large number of unreliable parts is challenging.  

Two widely used approaches for handling hardware-related reliability
issues are those based on \emph{checkpointing}, and on
\emph{functional replication}. The checkpoint-restore paradigm
requires the running application to periodically save its state on
non-volatile storage (e.g., disk) so that it can resume execution from
the last saved snapshot in case of failure. It should be observed that
saving a snapshot may require considerable time; therefore, the
interval between checkpoints must be carefully tuned to minimize the
overhead.

Functional replication consists on replicating parts of the
application on different execution nodes, so that failures can be
tolerated if there is some minimum number of running instances of each
component. Note that each component must be modified so that it is
made aware that multiple copies of its peers exist, and can interact
with all instances appropriately.

It is important to remark that functional replication is not effective
against logical errors, i.e., bugs in the running applications, since
the bug can be triggered at the same time on all instances. A
prominent -- and frequently mentioned -- example is the failure of the
Ariane 5 rocket that was caused by a software error on
its~\acp{IRP}. There were two~\ac{IRP}, providing hardware
fault-tolerance, but both used the same software. When the two
software instances were fed with the same (correct) input from the
hardware, the bug (an uncatched data conversion exception) caused both
programs to crash, leaving the rocket without
guidance~\cite{ariane-5}. The $N$-version programming
technique~\cite{n-version} can be used to protect against software
errors, and requires running several functionally equivalent programs
that have been independently developed from the same specifications.

In this paper, we present FT-GAIA, a fault-tolerant extension of the
GAIA/ART\`IS parallel and distributed simulation
middleware~\cite{gda-dsrt-2004, artis}. FT-GAIA is based on functional
replication, and can handle crash errors and byzantine faults, using
the concept of \emph{server groups}~\cite{cristian93}: simulation
entities are replicated so that the model can be executed even if some
of them fail. We show how functional replication can be implemented as
an additional software layer in the GAIA/ART\`IS stack; all
modifications are transparent to user-level simulation models,
therefore FT-GAIA can be used as a drop-in replacement to GAIA/ART\`IS
when fault tolerance is the major concern.

This paper is organized as follows. In Section~\ref{sec:related-work}
we review the art related to fault tolerance in~\ac{PADS}. The
GAIA/ART\`IS parallel and distributed simulation middleware is
described in Section~\ref{sec:gaia-artis}.
Section~\ref{sec:fault-tolerant-simulation} is devoted to the
description of FT-GAIA, a fault-tolerant extension to GAIA/ART\`IS.
An empirical performance evaluation of FT-GAIA, based on a prototype
implementation we have developed, is discussed in
Section~\ref{sec:experimental-evaluation}. Finally,
Section~\ref{sec:conclusions} provides some concluding remarks.

\section{Related Work}\label{sec:related-work}

Although fault tolerance is an important and widely discussed topic in
the context of distributed systems research, it received comparatively
little attention by the~\ac{PADS} community. The proposed approaches
for bringing fault tolerance to~\ac{PADS} are either based on
checkpointing or on functional replication, with a few works
considering also partially centralized architectures.

\subsection{Checkpointing}

In~\cite{Damani:1998:FDS:278008.278014} the authors propose a rollback
based optimistic recovery scheme in which checkpoints are periodically
saved on stable storage. The distributed simulation uses an optimistic
synchronization scheme, where out-of-order (``straggler'') events
cause rollbacks that are handled according to the Time Warp
protocol~\cite{Jefferson85}. The novel idea is to model failures as
straggler events with a timestamp equal to the last saved
checkpoint. In this way, the authors can leverage the Time Warp
protocol to handle failures.

In~\cite{Eklof:2005:FFH:1162708.1162915,Eklof:2006:EFM:1136644.1136877}
the authors propose a new framework called Distributed Resource
Management System (DRMS) to implement reliable IEEE 1516
federation~\cite{HLA}. The DRMS handles crash failures using
checkpoints saved to stable storage, that is then used to migrate
federates from a faulty host to a new host when necessary. The
simulation engine is again based on an optimistic synchronization
scheme, and the migration of federates is implemented through Web
services.

In~\cite{Chen20081487} the authors propose a decoupled federate
architecture in which each IEEE 1516 federate is separated into a
virtual federate process and a physical federate process. The former
executes the simulation model and the latter provides middleware
services at the backend. This solution enables the implementation of
fault-tolerant distributed simulation schemes through migration of
virtual federates.

The~CUMULVS middleware~\cite{Kohl:1998:EFF:281035.281042} introduces
the support for fault tolerance and migration of simulations based on
checkpointing. The middleware is not designed to support~\ac{PADS} but
it allows the migration of running tasks for load balancing and to
improve a task's locality with a required resource.

A slightly different approach is proposed in~\cite{Luthi2004}. In 
which, the authors introduce the Fault Tolerant Resource Sharing System 
(FT-RSS) framework. The goal of FT-RSS is to build fault tolerant 
IEEE 1516 federations using an architecture in which a separate FTP
server is used as a persistent storage system. The persistent storage
is used to implement the migration of federates from one node to another.
The FT-RSS middleware supports replication of federates, partial
failures and fail-stop failures.

\subsection{Functional Replication}

In~\cite{Agrawal:1992:ROT:167293.167662} the authors propose the use
of functional replication in Time Warp simulations with the aim to
increase the simulator performance and to add fault
tolerance. Specifically, the idea is to have copies of the most
frequently used simulation entities at multiple sites with the aim of
reducing message traffic and communication delay. This approach is
used to build an optimistic fault tolerance scheme in which it is
assumed that the objects are fault free most of the time. The rollback
capabilities of Time Warp are then used to correct intermittent and
permanent faults.

In~\cite{Liris-4840} the authors describe DARX, an adaptive
replication mechanism for building reliable multi-agent systems. Being
targeted to multi-agent systems, rather than~\ac{PADS}, DARX is mostly
concerned with adaptability: agents may change their behavior at any
time, and new agents may join or leave the system. Therefore, DARX
tries to dynamically identify which agents are more ``important'', and
what degree of replication should be used for those agents in order to
achieve the desired level of fault-tolerance. It should be observed
that DARX only handles crash failures, while FT-GAIA also deals with
Byzantine faults.

\section{The GAIA-ART\`IS Middleware}\label{sec:gaia-artis}

To make this paper self-contained, we provide in this section a brief
introduction of the GAIA/ART\`IS parallel and distributed simulation
middleware; the interested reader is referred to~\cite{gda-dsrt-2004, artis,
  gda-simpat-2014} and the software homepage~\cite{pads}.

The \textit{Advanced RTI System} (ART\`IS) is a parallel and
distributed simulation middleware loosely inspired by the Runtime
Infrastructure described in the IEEE~1516 standard ``High Level
Architecture'' (HLA)~\cite{ieee1516}.  ART\`IS implements a
parallel/distributed architectures where the simulation model is
partitioned in a set of~\acp{LP}~\cite{Fuj00}. As described in
Section~\ref{sec:introduction}, the execution architecture in charge
of running the simulation is composed of interconnected~\acp{PE} and
each~\ac{PE} runs one or more~\acp{LP} (usually, a~\ac{PE} hosts
one~\ac{LP}).

In a~\acp{PADS}, the interactions between the model components are
driven by message exchanges. The low computation/communication ratio
makes~\ac{PADS} communication-bound, so that the wall-clock execution
time of distributed simulations is highly dependent on the performance
of the communication network (i.e., latency, bandwidth and
jitter). Reducing the communication overhead can be crucial to speed
up the event processing rate of~\ac{PADS}. This can be achieved by
clustering interacting entities on the same physical host, so that
communications can happen through shared memory. 

Among the various services provided by ART\`IS, time management (i.e.,
synchronization) is fundamental for obtaining correct simulation runs
that respect the causality dependencies of events. ART\`IS supports
both conservative (Chandy-Misra-Bryant~\cite{cmb}) and optimistic
(Time Warp~\cite{Jefferson85}) synchronization algorithms. Moreover, a
very simple time-stepped synchronization is supported.

The \textit{Generic Adaptive Interaction Architecture} (GAIA) is a
software layer built on top of ART\`IS~\cite{pads}. In~GAIA,
each~\ac{LP} acts as the container of some~\acp{SE}: the simulation
model is partitioned in its basic components (the~\acp{SE}) that are
allocated among the~\acp{LP}. The system behavior is modeled by the
interactions among the~\acp{SE}; such interactions take the form of
timestamped messages that are exchanged among the entities. From the
user's point of view, a simulation model based on GAIA/ART\`S follows
a Multi Agent System (MAS) approach. In fact, each~\ac{SE} is an
autonomous agent that performs some actions (individual behavior) and
interacts with other agents in the simulation.

In most cases, the interaction between the~\acp{SE} of a~\ac{PADS} are
not completely uniform, meaning that there are clusters of~\acp{SE}
where internal interactions are more frequent. The structure of these
clusters of highly interacting entities may change over time, as the
simulation model evolves. The identification of such clusters is
important to improve the performance of a~\ac{PADS}: indeed, by
putting heavily-interacting entities on as few~\acp{LP} as possible,
we may replace most of the expensive LAN/WAN communications by more
efficient shared memory messages.

In GAIA, the analysis of the communication pattern is based on simple
self-clustering heuristics~\cite{gda-simpat-2014}. For example, in
the default heuristic, every few timesteps for each~\ac{SE} is found
which~\ac{LP} is the destination of the large percentage
of interactions. If it is not the~\ac{LP} in which the~\ac{SE} is
contained then a migration is triggered. The migration of~\acp{SE}
among~\acp{LP} is transparent to the simulation model developer; 
entities migration is useful not only to reduce the communication
overhead, but also to achieve better load-balancing among
the~\acp{LP}, especially on heterogeneous execution platforms where
execution units are not identical. In these cases, GAIA can migrate
entities away from less powerful~\acp{PE}, towards more capable
processors if available.

\section{Fault-Tolerant Simulation}\label{sec:fault-tolerant-simulation}

FT-GAIA is a fault-tolerant extension to the GAIA/ART\`IS distributed
simulation middleware. As will be explained below, FT-GAIA uses
functional replication of simulation entities to achieve tolerance
against crashes and Byzantine failures of the~\acp{PE}.

FT-GAIA is implemented as a software layer on top of GAIA and provides
the same functionalities of GAIA with only minor additions. Therefore,
FT-GAIA is mostly transparent to the user, meaning that any simulation
model built for GAIA can be easily ported to FT-GAIA.

\begin{figure}[t]
  \centering%
  \includegraphics[scale=.8]{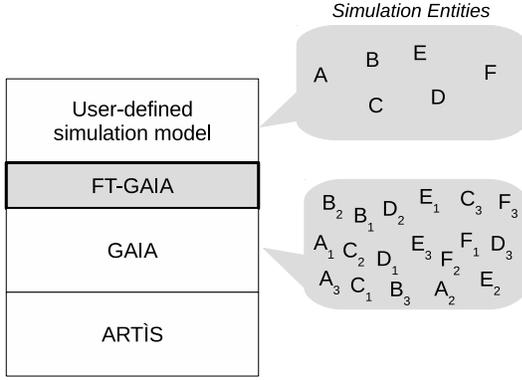}
  \caption{Layered structure of the FT-GAIA simulation engine. The
    user-defined simulation model defines a set of entities $\{A, B,
    C, D, E, F\}$; FT-GAIA creates multiple (in this example, 3)
    instances of each entity, that are handled by
    GAIA.}\label{fig:gaiaft}
\end{figure}

FT-GAIA works by replicating simulation entities (see
Fig.~\ref{fig:gaiaft}) to tolerate crash-failures and byzantine faults
of the~\acp{PE}. A crash may be caused by a failure of the hardware --
including the network connection -- and operating system. A byzantine
failure refers to an arbitrary behavior of a~\ac{PE} that causes
the~\ac{LP} to crash, terminate abnormally, or to send arbitrary
messages (including no messages at all) to other~\acp{PE}.

Replication is based on the following principle. If a conventional,
non-fault tolerant distributed simulation is composed of $N$ distinct
simulation entities, FT-GAIA generates $N \times M$ entities, by
generating $M$ independent instances of each simulation entity. All
instances $A_1, \ldots A_M$ of the same entity $A$ perform the same
computation: if no fault occurs, they produce the same result. 

Replication comes with a cost, both in term of additional processing
power that is needed to execute all instances, and also in term of an
increased communication load between the~\acp{LP}. Indeed, if two
entities $A$ and $B$ communicate by sending a message from $A$ to $B$,
then after replication each instance $A_i$ must send the same message
to all instances $B_j$, $1 \leq i, j \leq M$, resulting in $M^2$
(redundant) messages. Therefore, the level of replication $M$ must be
chosen wisely in order to achieve a good balance between overhead and
fault tolerance, also depending on the types of failures (crash
failures or Byzantine faults) that the user wants to address.

\paragraph*{Handling crash failures} A crash failure happens
when a~\ac{PE} halts, but operated correctly until it halted. In this
case, all simulation entities running on that~\ac{PE} stop their
execution and the local state of computation is lost. From the theory
of distributed systems, it is known that in order to tolerate $f$
crash failures we must execute at least $M = f+1$ instances of each
simulation entity. Each instance must be executed on a
different~\acp{PE}, so that the failure of a~\ac{PE} only affects one
instance of all entities executed there. This is is equivalent to
running $M$ copies of a monolithic (sequential) simulation, with the
difference that a sequential simulation does not incur in
communication and synchronization overhead. However, unlike sequential
simulations, FT-GAIA can take advantage of more than $M$ \acp{PE}, by
distributing all $N \times M$ entities on the available execution
units. This reduces the workload on the~\acp{PE}, reducing the
wall-clock execution time of the simulation model.

\paragraph*{Handling Byzantine Failures}
Byzantine failures include all types of abnormal behaviors of
a~\ac{PE}. Examples are: the crash of a component of the distributed
simulator (e.g., \ac{LP} or entity); the transmission of
erroneous/corrupted data from an entity to other entities; computation
errors that lead to erroneous results. In this case $M = 2f+1$
replicas of a system are needed to tolerate up to $f$ byzantine faults
in a distributed system using the ``majority'' rule: an~\ac{SE}
instance $B_i$ can process an incoming message $m$ from $A_j$ when it
receives at least $f+1$ copies of $m$ from different instances of the
sender entity $A$. Again, all $M$ instances of the same~\ac{SE} must
be located on different~\acp{PE}.

\paragraph*{Allocation of Simulation Entities}
Once the level of replication $M$ has been set, it is necessary to
decide where to create the $M$ instances of all~\acp{SE}, so that the
constraint that each instance is located on a different~\ac{PE} is
met. In FT-GAIA the deployment of instances is performed during the
setup of the simulation model. In the current implementation, there is
a centralized service that keeps track of the initial location of
all~\ac{SE} instances.  When a new~\ac{SE} is created, the service
creates the appropriate number of instances according to the
redundancy model to be employed, and assigns them to the~\acp{LP} so
that all instances are located on different~\acp{LP}. Note that all
instances of the same~\ac{SE} receive the same initial seed for their
internal pseudo-random number generators; this guarantees that their
execution traces are the same, regardless of the~\ac{LP} where
execution occurs and the degree of replication.

\paragraph*{Message Handling}
We have already stated that fault-tolerance through functional
replication has a cost in term of increased message load
among~\acp{SE}. Indeed, for a replication level $M$ (i.e., there are
$M$ instances of each~\ac{SE}) the number of messages exchanged
between entities grows by a factor of $M^2$.

A consequence of message redundancy is that message filtering must be
performed to avoid that multiple copies of the same message are
processed more than once by the same~\ac{SE} instance. FT-GAIA takes
care of automatically filtering the excess messages according to the
fault model adopted; filtering is done outside of the~\ac{SE}, which
are therefore totally unaware of this step. In the case of crash
failures, only the first copy of each message that is received by
a~\ac{SE} is processed; all further copies are dropped by the
receiver. In the case of Byzantine failures with replication level
$M=2f+1$, each entity must wait for at least $f+1$ copies of the same
message before it can handle it. Once a strict majority has been
reached, the message can be processed and all further copies of the
same messages that might arrive later on can be dropped.

\paragraph*{Entities Migration}
\ac{PADS} can benefit from migration of entities to balance
computation/communication load and reduce the communication cost, by
placing entities that interact frequently ``next'' to each other
(e.g., on the same~\ac{LP})~\cite{gda-simpat-2014}. In FT-GAIA, entity
migration is subject to the constraint that instances of the
same~\ac{SE} can never reside on the same~\ac{LP}. Entity migration is
handled by the underlying GAIA/ART\`IS
middleware~\cite{gda-dsrt-2004}: each~\ac{LP} runs a fully distributed
``clustering heuristic'' that tries to put together (i.e., on the
same~\ac{LP}) the~\acp{SE} that interact frequently through message
exchanges. Special care is taken to avoid putting too many entities on
the same~\acp{LP} that would become a bottleneck. Once a new feasible
allocation is found, the entities are migrated by moving their state
to the new~\ac{LP}.

\section{Experimental Evaluation}\label{sec:experimental-evaluation}

In this section we evaluate a prototype implementation of FT-GAIA by
implementing a simple simulation model of a Peer-to-Peer communication
system. We execute the simulation model with FT-GAIA under different
workload parameters (described below) and record the~\ac{WCT}
(excluding the time to setup the simulation) and other metrics of
interest. The tests were performed on a cluster of workstations, each
being equipped with an Intel Core i5-4590 3.30 GHz processors with 8
GB of RAM. The Operating System was Debian Jessie. The workstations
are connected through a Fast Ethernet LAN.

\subsection{Simulation Model}

We simulate a simple P2P communication protocol over randomly
generated directed overlay graphs. Nodes of the graphs are peers,
while links represent communication
connections~\cite{D'Angelo:2009,gda-mospas-11}.  In these overlays,
nodes have all the same out-degree, that has been set to $5$ in our
experiments.  During the simulation, each node periodically updates
its neighbor set.  Latencies for message transmission over overlay
links are generated using a lognormal distribution~\cite{Farber:2002}.

The simulated communication protocol works as follows. Periodically,
nodes send PING messages to other nodes, that in turn reply with a
PONG message that is used by the sender to estimate the average
latencies of the links (note that communication links are, in fact,
bidirectional). The destination of a PING is randomly selected to be a
neighbor (with probability $p$), or a non-neighbor (with probability
$1-p$). A neighbor is a node that can be reached through an outgoing
link in the directed overlay graph.

Each node of the P2P overlay is represented by a~\ac{SE} within
some~\ac{LP}. Unless stated otherwise, each~\ac{LP} was executed on a
different~\ac{PE}, so that no two~\acp{LP} shared their execution
node. We consider three scenarios: a \emph{no fault} scenario, where
no faults occur, a \emph{crash} scenario, where crash failures occurs,
and a \emph{Byzantine} scenario where Byzantine faults occurs.

We executed 15 independent replications of each simulation run. In all
the following charts, mean values are reported with a 99.5\%
confidence interval.

\subsection{Impact of the number of~\acp{LP} and ~\acp{SE}}

\begin{figure}[t]
  \centering\includegraphics[width=\columnwidth]{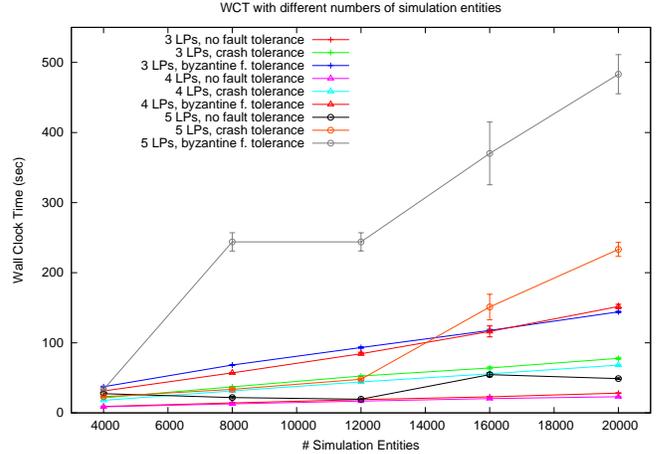}
  \caption{\acl{WCT} as a function of the number of~\acp{LP}, for
    varying number of~\acp{SE}. The number of~\acp{LP} is equal to the
    number of~\acp{PE}. Migration is disabled. Lower is better.}\label{fig:diffENT}
\end{figure}

Figure~\ref{fig:diffENT} shows the~\ac{WCT} of the simulation that was
executed for 10000 timesteps with a varying number of~\acp{SE}; recall
that the number of~\acp{SE} is equal to the number of nodes in the P2P
overlay graph. The number of~\acp{LP} was set to 3, 4, and 5. We show
the~\ac{WCT} for the three failure scenarios we are considering: no
failure, a single crash, and a single Byzantine failure. In all these
cases the self-clustering (i.e.~migration) is disabled.

Results with~3 and~4 \acp{LP} are similar, with a slight improvement
with 4~\acp{LP}. Conversely, higher~\ac{WCT} is observed when
5~\acp{LP} are used.  As expected, the higher the number of~\acp{SE}
the higher the~\ac{WCT}, since the simulation incurs in a higher
communication overhead. Moreover, all curves have a similar trend. In
particular, the increment due to the faults management schemes is
mainly due to the higher amount of messages exchanged among nodes.

\begin{figure}[t]
  \centering%
  \includegraphics[width=\columnwidth]{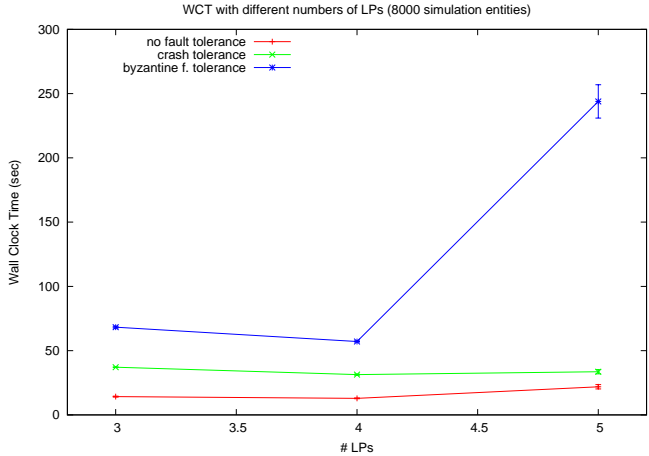}
  \caption{\acl{WCT} as a function of the number of~\acp{LP}, with
    8000~\acp{SE}. Migration is disabled. Lower is better.}\label{fig:diffLP345_8k}
\end{figure}

\begin{figure}[t]
  \centering%
  \includegraphics[width=\columnwidth]{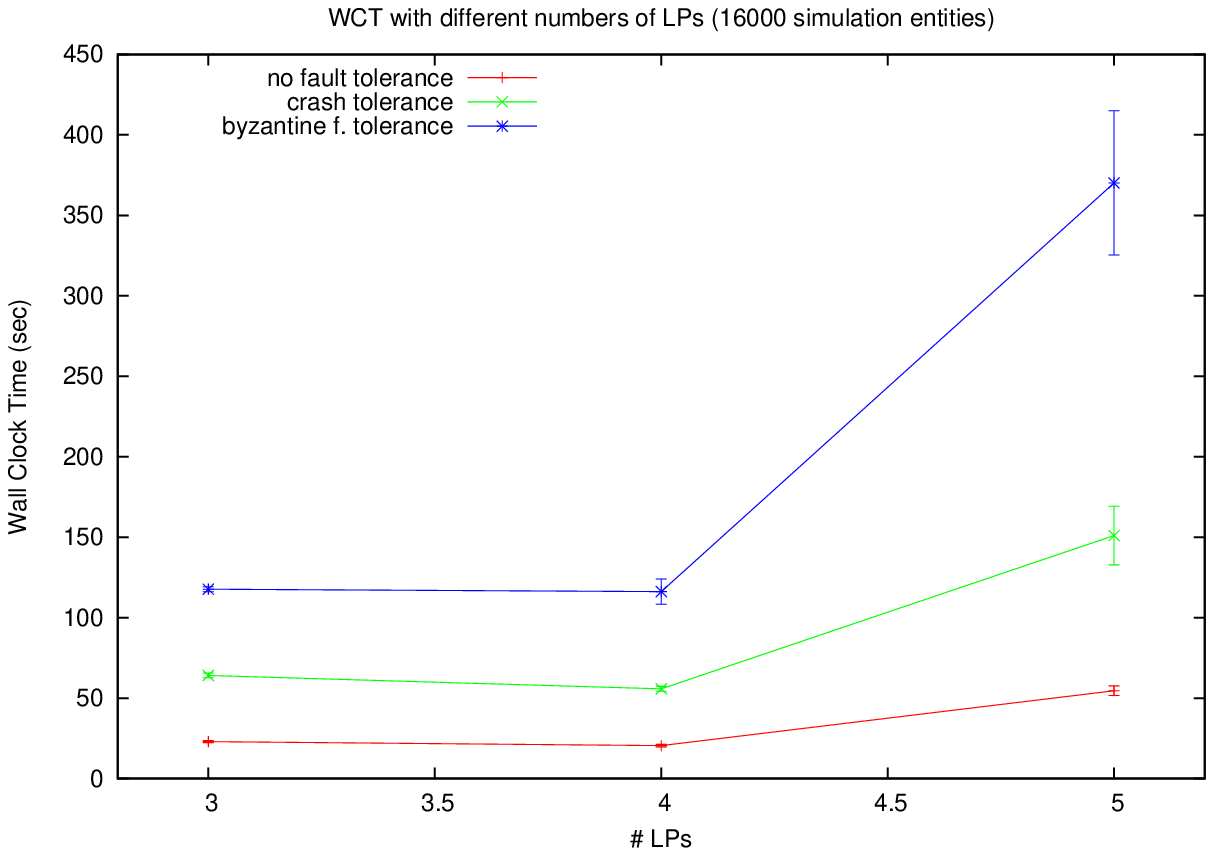}
  \caption{\ac{WCT} as a function of the number of~\acp{LP}, with
    10000~\acp{SE}. Migration is disabled. Lower is better.}\label{fig:diffLP345_16k}
\end{figure}

Figures~\ref{fig:diffLP345_8k} and~\ref{fig:diffLP345_16k} show
the~\ac{WCT} when varying the number of LPs, with~8000
and~16000~\acp{SE}, respectively.
The two charts emphasize the increment of the time required to
terminate the simulations with 5~\acp{LP} and in presence of Byzantine
faults. This is due to the increased number of messages exchanged
among the~\acp{LP}: each message needs to be sent to three ($2M+1$) different
destinations in order to guarantee fault tolerance.

\subsection{Impact of the number of LPs per host}

\begin{figure}[t]
  \centering%
  \includegraphics[width=\columnwidth]{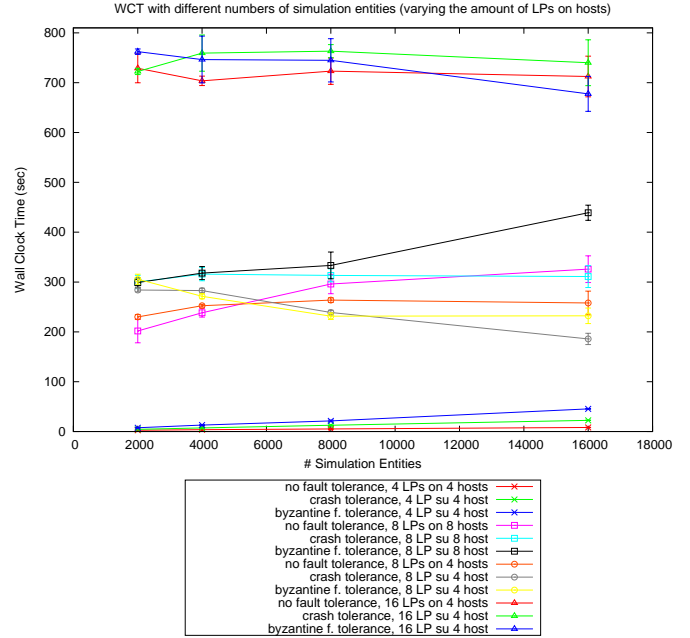}
  \caption{\ac{WCT} as a function of the number of~\acp{LP}, with
    different numbers of~\acp{LP} for each~\ac{PE}. Migration is disabled. Lower is
    better.}\label{fig:diffENT_488b16q}
\end{figure}

In the previous experiments, we placed each~\ac{LP} in a
different~\ac{PE}. Figure~\ref{fig:diffENT_488b16q} shows the~\ac{WCT}
when more than one~\ac{LP} is placed in a~\ac{PE}. In particular, we
consider the following scenarios: (\emph{i}) 4~\acp{LP} placed over
4~\acp{PE} (1~\ac{LP} per host), (\emph{ii}) 8~\acp{LP} placed over
8~\acp{PE} (1~\ac{LP} per host), (\emph{iii}) 8~\acp{LP} placed over
4~\acp{PE} (2~\acp{LP} per host), and (\emph{iv}) 16~\acp{LP} over
4~\acp{PE} (4~\acp{LP} per host).  For each scenario, we consider the
three failure scenarios already mentioned (no failures, crash,
Byzantine failures). Also in these cases, the migration is disabled.
Each curve in the figure is related to one of those scenarios, when
varying the amount of~\acp{SE}. It is worth noting that, when two or
more~\acp{LP} are run on the same~\ac{PE}, they can communicate using
shared memory rather than by LAN.

We observe that the scenario with 4~\acp{LP} over 4~\acp{PE} is
influenced by the number of~\acp{SE} and the failure scenario, while in
the other cases it is the number of~\acp{LP} that mainly determines
the simulator performance.  When 8~\acp{LP} are present,
slightly better results are obtained with 4~\acp{LP} (rather than 8).
This is due to the better communication efficiency (e.g.~reduced
latency) provided by the shared memory with the respect to the LAN
protocols.

The worst performance is measured when 16~\acp{LP} are executed on
4~\acp{PE}. This is due to the fact that the amount of computation
in the simulation model is quite limited. Therefore, partitioning
the~\acp{SE} in 16~\acp{LP} has the effect to increase the
communication cost without any benefit under the computational
point of view (i.e.~in the model there is not enough computation
to be parallelized).

\subsection{Impact of the number of failures}

\begin{figure}[t]
  \centering%
  \includegraphics[width=\columnwidth]{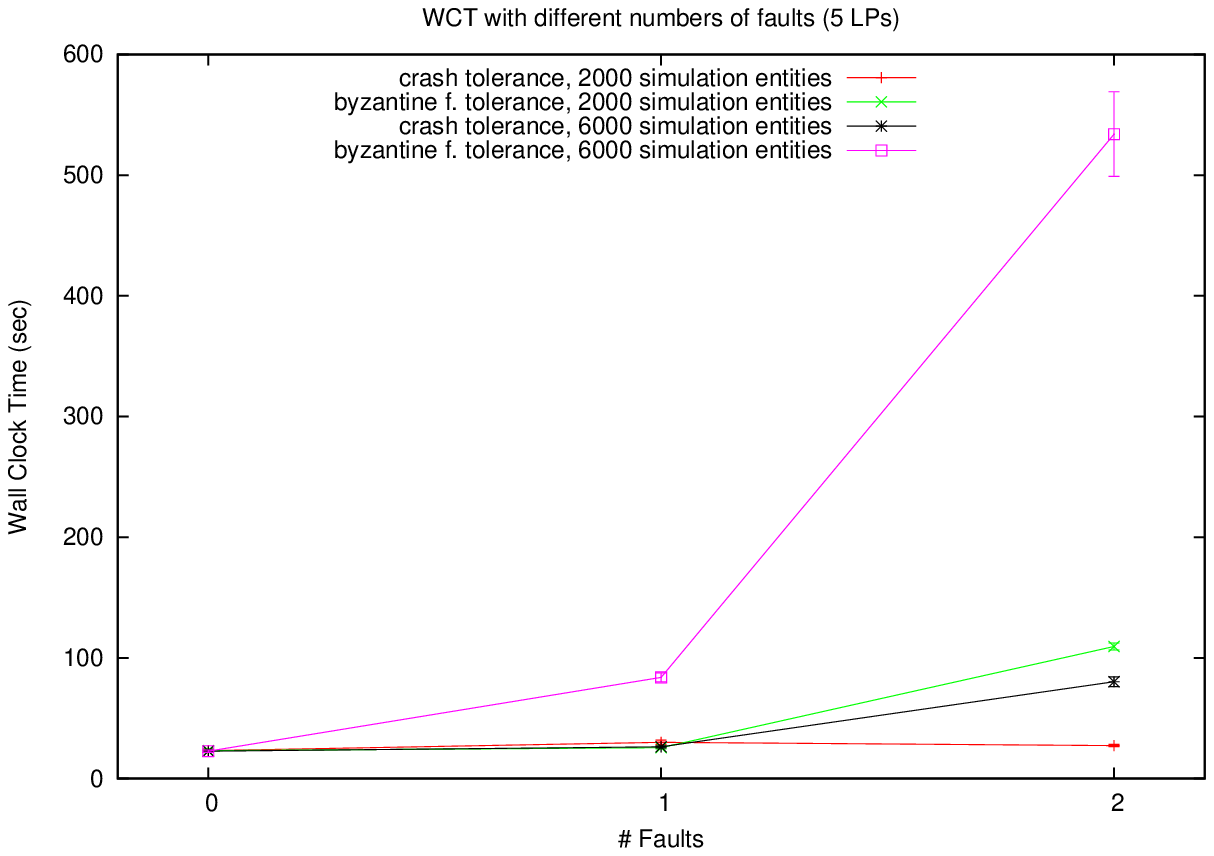}
  \caption{\ac{WCT} as a function of the number of faults; 10000
    timesteps with 5~\acp{LP}. Migration is disabled. Lower is
    better.}\label{fig:diffMAXFT012}
\end{figure}

\begin{figure}[t]
  \centering%
  \includegraphics[width=\columnwidth]{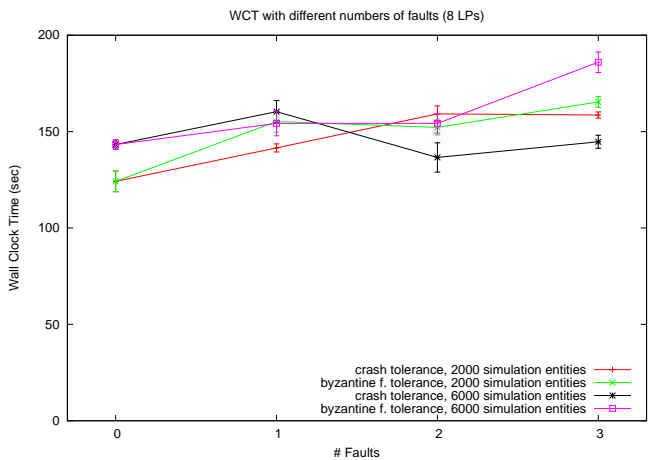}
  \caption{\ac{WCT} as a function of the number of faults; 2000
    timesteps over 8 LPs. Migration is disabled. Lower is better.}\label{fig:diffMAXFT0123}
\end{figure}

We now study the impact of the number of faults on the
simulation~\ac{WCT}. We consider two scenarios, one with 5~\acp{LP}
over 5~\acp{PE} (Figure~\ref{fig:diffMAXFT012}), and one with
8~\acp{LP} over 4~\acp{PE} (Figure \ref{fig:diffMAXFT0123}).  The
choice of 5~\acp{LP} is motivated by the fact that this is the minimum
number of~\acp{LP} that allows us to tolerate up to 2 Byzantine
faults. 
The scenario with 8~\acp{LP} on 4~\acp{PE} allows testing 3 Byzantine
faults with 2~\acp{LP} per hosts, reducing the communication overhead.

Figure~\ref{fig:diffMAXFT012} shows the WCTs measured with 0, 1 and 2
faults. Each curve refers to a scenario composed of 2000 or 6000 SEs
with crash or Byzantine failures. As expected, the higher the number
of faults, the higher the~\acp{WCT}, especially when Byzantine faults
are considered. Indeed, in this case a higher amount of communication
messages is required among nodes in order to properly handle the
faults.

A higher~\ac{WCT} is measured with 8~\acp{LP}, as shown in
Figure~\ref{fig:diffMAXFT0123}. In this case, the amount of faults
does not influence the simulation performance too much.
As before, the computational load of this simulation model is too low
for gaining from the partitioning in 8~\acp{LP}. 
In other words, the latency introduced by the network communications is so high
that both the number of SEs and and the number of faults have a negligible impact.

\subsection{Impact of~\acp{SE} migration}

\begin{figure}[!htp]
  \centering%
  \includegraphics[width=\columnwidth]{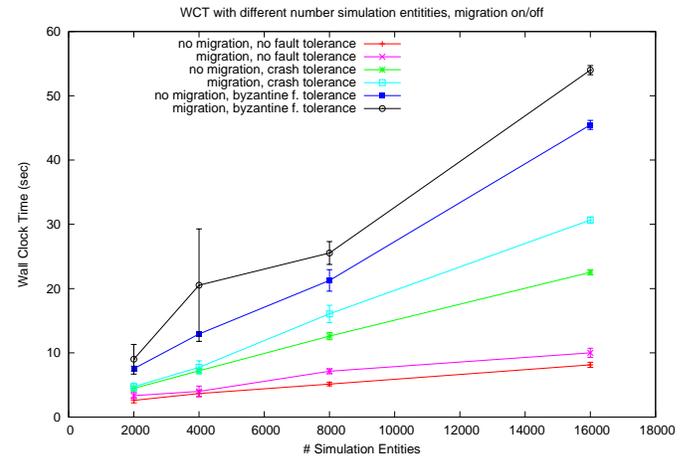}
  \caption{\ac{WCT} with~\acp{SE} migration ON/OFF, as a function of the
    number of~\acp{SE}. Lower is better.}\label{fig:diffMIGR}
\end{figure}

Figure~\ref{fig:diffMIGR} shows the~\ac{WCT} with different failure
schemes, when~\acp{SE} migration is enabled/disabled.  In this case,
the trend obtained with the~\acp{SE} migration is similar to that
obtained when no migration is performed but the overall performance
are better when the migration is turned off. This is due to the
overhead introduced by the self-clustering heuristics and the
~\acp{SE} state that is transfered between the~\acp{LP}. In other
words, the adaptive clustering of~\acp{SE}, in this case, is 
unable to give a speedup.

It is worth noting that, in this prototype, we have decided to
use the very general clustering heuristics that were already 
implemented in GAIA/ART\`S. We think that, more more specific
heuristics will be able to improve the clustering performance
and therefore balance the overhead introduced by the support of 
fault tolerance.

\section{Conclusions and Future Work}\label{sec:conclusions}

In this paper we described FT-GAIA, a software-based fault-tolerant
extension of the GAIA/ART\`IS parallel and distributed simulation
middleware. FT-GAIA transparently replicates simulation entities and
distributes them on multiple execution nodes. In this way, the
simulation can tolerate crash-failures and Byzantine faults of
computing nodes.  FT-GAIA can benefit from the automatic load
balancing facilities provided by GAIA/ART\`IS that allow simulated
entities to be migrated among execution nodes. A preliminary
performance evaluation of FT-GAIA has been presented, based on a
prototype implementation. Results show that a high degree of fault
tolerance can be achieved, at the cost of a moderate increase in the
computational load of the execution units.

As a future work, we aim at improving the efficiency of FT-GAIA by
leveraging on ad-hoc clustering heuristics. Indeed, we believe that
specifically tuned clustering and load balancing mechanisms can
significantly reduce the overhead introduced by the replication of the
simulated entities.

\section*{Acronyms}

\begin{acronym}[PDES]
  \acro{DES}{Discrete Event Simulation}
  \acro{FEL}{Future Event List}
  \acro{GVT}{Global Virtual Time}
  \acro{IRP}{Inertial Reference Platform}
  \acro{LVT}{Local Virtual Time}
  \acro{LP}{Logical Process}
  \acroplural{LPs}{Logical Processes}
  \acro{MTTF}{Mean Time To Failure}
  \acro{PADS}{Parallel and Distributed Simulation}
  \acro{PE}{Processing Element}
  \acro{SE}{Simulated Entity}
  \acroplural{SEs}{Simulated Entities}
  \acro{WCT}{Wall Clock Time}
\end{acronym}

\bibliographystyle{IEEEtran}
\bibliography{IEEEabrv,ftgaia}

\begin{thebibliography}{10}
\providecommand{\url}[1]{#1}
\csname url@samestyle\endcsname
\providecommand{\newblock}{\relax}
\providecommand{\bibinfo}[2]{#2}
\providecommand{\BIBentrySTDinterwordspacing}{\spaceskip=0pt\relax}
\providecommand{\BIBentryALTinterwordstretchfactor}{4}
\providecommand{\BIBentryALTinterwordspacing}{\spaceskip=\fontdimen2\font plus
\BIBentryALTinterwordstretchfactor\fontdimen3\font minus
  \fontdimen4\font\relax}
\providecommand{\BIBforeignlanguage}[2]{{%
\expandafter\ifx\csname l@#1\endcsname\relax
\typeout{** WARNING: IEEEtran.bst: No hyphenation pattern has been}%
\typeout{** loaded for the language `#1'. Using the pattern for}%
\typeout{** the default language instead.}%
\else
\language=\csname l@#1\endcsname
\fi
#2}}
\providecommand{\BIBdecl}{\relax}
\BIBdecl

\bibitem{Fuj00}
R.~M. Fujimoto, \emph{Parallel and distributed simulation systems}, ser. Wiley
  series on parallel and distributed computing.\hskip 1em plus 0.5em minus
  0.4em\relax Wiley, 2000.

\bibitem{reliability-wall}
X.~Yang, Z.~Wang, J.~Xue, and Y.~Zhou, ``The reliability wall for exascale
  supercomputing,'' \emph{Computers, IEEE Transactions on}, vol.~61, no.~6, pp.
  767--779, 2012.

\bibitem{bolch}
G.~Bolch, S.~Greiner, H.~de~Meer, and K.~Trivedi, \emph{Queueing Networks and
  Markov Chains: Modeling and Performance Evaluation with Computer Science
  Applications}.\hskip 1em plus 0.5em minus 0.4em\relax Wiley, 1998.

\bibitem{ariane-5}
M.~Dowson, ``The ariane 5 software failure,'' \emph{SIGSOFT Softw. Eng. Notes},
  vol.~22, no.~2, pp. 84--, Mar. 1997.

\bibitem{n-version}
A.~Avizienis, ``The {N}-version approach to fault-tolerant software,''
  \emph{IEEE Trans. Softw. Eng.}, vol.~11, no.~12, pp. 1491--1501, Dec. 1985.

\bibitem{gda-dsrt-2004}
L.~Bononi, M.~Bracuto, G.~D'Angelo, and L.~Donatiello, ``A new adaptive
  middleware for parallel and distributed simulation of dynamically interacting
  systems,'' in \emph{Proceedings of the 8th IEEE International Symposium on
  Distributed Simulation and Real-Time Applications}.\hskip 1em plus 0.5em
  minus 0.4em\relax Washington, DC, USA: IEEE Computer Society, 2004, pp.
  178--187.

\bibitem{artis}
------, ``{ART\`{I}S}: A parallel and distributed simulation middleware for
  performance evaluation,'' in \emph{ISCIS}, ser. Lecture Notes in Computer
  Science, C.~Aykanat, T.~Dayar, and I.~Korpeoglu, Eds., vol. 3280.\hskip 1em
  plus 0.5em minus 0.4em\relax Springer, 2004, pp. 627--637.

\bibitem{cristian93}
F.~Cristian, ``Understanding fault-tolerant distributed systems,''
  \emph{Commun. ACM}, vol.~34, no.~2, pp. 56--78, Feb. 1991.

\bibitem{Damani:1998:FDS:278008.278014}
O.~P. Damani and V.~K. Garg, ``Fault-tolerant distributed simulation,'' in
  \emph{Proceedings of the Twelfth Workshop on Parallel and Distributed
  Simulation}, ser. PADS '98.\hskip 1em plus 0.5em minus 0.4em\relax
  Washington, DC, USA: IEEE Computer Society, 1998, pp. 38--45.

\bibitem{Jefferson85}
D.~R. Jefferson, ``Virtual time,'' \emph{ACM Trans. Program. Lang. Syst.},
  vol.~7, no.~3, pp. 404--425, Jul. 1985.

\bibitem{Eklof:2005:FFH:1162708.1162915}
M.~Ekl\"{o}f, F.~Moradi, and R.~Ayani, ``A framework for fault-tolerance in
  hla-based distributed simulations,'' in \emph{Proceedings of the 37th
  Conference on Winter Simulation}, ser. WSC '05.\hskip 1em plus 0.5em minus
  0.4em\relax Winter Simulation Conference, 2005, pp. 1182--1189.

\bibitem{Eklof:2006:EFM:1136644.1136877}
M.~Eklof, R.~Ayani, and F.~Moradi, ``Evaluation of a fault-tolerance mechanism
  for hla-based distributed simulations,'' in \emph{Proceedings of the 20th
  Workshop on Principles of Advanced and Distributed Simulation}, ser. PADS
  '06.\hskip 1em plus 0.5em minus 0.4em\relax Washington, DC, USA: IEEE
  Computer Society, 2006, pp. 175--182.

\bibitem{HLA}
``{IEEE Standard for Modeling and Simulation (M\&S) High Level Architecture
  (HLA)--Framework and Rules},'' {IEEE Std 1516-2010 (Revision of IEEE Std
  1516-2000)}, pp. 1--38, 2010.

\bibitem{Chen20081487}
D.~Chen, S.~J. Turner, W.~Cai, and M.~Xiong, ``A decoupled federate
  architecture for high level architecture-based distributed simulation,''
  \emph{Journal of Parallel and Distributed Computing}, vol.~68, no.~11, pp.
  1487 -- 1503, 2008.

\bibitem{Kohl:1998:EFF:281035.281042}
J.~A. Kohl and P.~M. Papadopoulas, ``Efficient and flexible fault tolerance and
  migration of scientific simulations using cumulvs,'' in \emph{Proceedings of
  the SIGMETRICS Symposium on Parallel and Distributed Tools}, ser. SPDT
  '98.\hskip 1em plus 0.5em minus 0.4em\relax New York, NY, USA: ACM, 1998, pp.
  60--71.

\bibitem{Luthi2004}
J.~L{\"u}thi and S.~Gro{\ss}mann, \emph{Computational Science - ICCS 2004: 4th
  International Conference, Krak{\'o}w, Poland, June 6-9, 2004, Proceedings,
  Part III}.\hskip 1em plus 0.5em minus 0.4em\relax Berlin, Heidelberg:
  Springer Berlin Heidelberg, 2004, ch. FT-RSS: A Flexible Framework for Fault
  Tolerant HLA Federations, pp. 865--872.

\bibitem{Agrawal:1992:ROT:167293.167662}
D.~Agrawal and J.~R. Agre, ``Replicated objects in time warp simulations,'' in
  \emph{Proceedings of the 24th Conference on Winter Simulation}, ser. WSC
  '92.\hskip 1em plus 0.5em minus 0.4em\relax New York, NY, USA: ACM, 1992, pp.
  657--664.

\bibitem{Liris-4840}
\BIBentryALTinterwordspacing
Z.~{Guessoum}, J.-P. {Briot}, N.~{Faci}, and O.~{Marin},
  ``\BIBforeignlanguage{en}{{Towards Reliable Multi-Agent Systems. An Adaptive
  Replication Mechanism }},'' \emph{\BIBforeignlanguage{en}{International
  Journal of MultiAgent and Grid Systems}}, vol.~6, no.~1, 2010. [Online].
  Available: \url{http://liris.cnrs.fr/publis/?id=4840}
\BIBentrySTDinterwordspacing

\bibitem{gda-simpat-2014}
G.~D'Angelo and M.~Marzolla, ``New trends in parallel and distributed
  simulation: From many-cores to cloud computing,'' \emph{Simulation Modelling
  Practice and Theory (SIMPAT)}, 2014.

\bibitem{pads}
``{Parallel And Distributed Simulation (PADS) research group},''
  \url{http://pads.cs.unibo.it}, 2016.

\bibitem{ieee1516}
{IEEE 1516 Standard, Modeling and Simulation ({M\&S}) High Level Architecture
  (HLA)}, 2000.

\bibitem{cmb}
K.~M. Chandy and J.~Misra, ``Asynchronous distributed simulation via a sequence
  of parallel computations,'' \emph{Commun. ACM}, vol.~24, no.~4, pp. 198--206,
  Apr. 1981.

\bibitem{D'Angelo:2009}
G.~D'Angelo and S.~Ferretti, ``Simulation of scale-free networks,'' in
  \emph{Proc. of International Conference on Simulation Tools and Techniques},
  ser. Simutools '09, 2009, pp. 20:1--20:10.

\bibitem{gda-mospas-11}
------, ``{LUNES: Agent-based Simulation of P2P Systems},'' in
  \emph{Proceedings of the International Workshop on Modeling and Simulation of
  Peer-to-Peer Architectures and Systems (MOSPAS 2011)}.\hskip 1em plus 0.5em
  minus 0.4em\relax IEEE, 2011.

\bibitem{Farber:2002}
J.~F\"{a}rber, ``Network game traffic modelling,'' in \emph{Proceedings of the
  1st Workshop on Network and System Support for Games}, ser. NetGames
  '02.\hskip 1em plus 0.5em minus 0.4em\relax New York, NY, USA: ACM, 2002, pp.
  53--57.

\end{thebibliography}

\end{document}